\documentclass[prl,twocolumn,superscriptaddress,showpacs,nobalancelastpage]{revtex4}

\usepackage{graphicx}

\begin{document}

\title{Efficient C-Phase gate for single-spin qubits in quantum dots}

\author{T. Meunier}
\affiliation{Kavli Institute of Nanoscience, Delft University of
Technology, PO Box 5046, 2600 GA Delft, Netherlands}
\affiliation{Institut N\'eel, CNRS and Universit\'e Joseph Fourier,38042 Grenoble, France}

\author{V. E. Calado}
\affiliation{Kavli Institute of Nanoscience, Delft University of
Technology, PO Box 5046, 2600 GA Delft, Netherlands}

\author{L. M. K. Vandersypen}
\affiliation{Kavli Institute of Nanoscience, Delft University of
Technology, PO Box 5046, 2600 GA Delft, Netherlands}

\date{\today}

\begin{abstract}

Two-qubit interactions are at the heart of quantum information processing. For single-spin qubits in semiconductor quantum dots, the exchange gate has always been considered the natural two-qubit gate. The recent integration of magnetic field or $g$-factor gradients in coupled quantum dot systems allows for a one-step, robust realization of the controlled phase (C-Phase) gate instead. We analyze the C-Phase gate durations and fidelities that can be obtained under realistic conditions, including the effects of charge and nuclear field fluctuations, and find gate error probabilities of below $10^{-4}$, possibly allowing fault-tolerant quantum computation.

\end{abstract}


\maketitle

The spin of a single electron is a natural degree of freedom for storing quantum information. In semiconductor quantum dots, all operations required for quantum computation with single spins have been demonstrated experimentally in recent years: single-shot read-out~\cite{NatureReadout} and coherent rotations of individual spins~\cite{KoppensNature} as well as coherent interactions between two neighbouring spins~\cite{PettaScience}. Single-spin relaxation times, $T_1$, are extremely long, up to 1 s ~\cite{NatureReadout,Amasha}. Recent measurements in GaAs quantum dots gave spin-echo coherence times, $T_{2,echo}$, of up to a few hundred $\mu s$~\cite{T2long}. In quantum dots built from materials without nuclear spins, even longer spin coherence times are expected.

Despite this impressive progress, there have been no experimental reports of elementary two-qubit gates such as the controlled-NOT gate or the C-Phase gate. These gates are important since they are, together with single-qubit rotations, the natural building blocks for quantum algorithms. Implementing the quantum Fourier transform for instance, comes down to concatenating a sequence of two-qubit C-Phase gates~\cite{Nielsen-Chuang}.

In their seminal paper, Loss and DiVincenzo~\cite{LossDiVincenzo} present a scheme to achieve a two-qubit C-Phase gate by two periods of coherent evolution under the exchange interaction between the spins, separated by a single qubit rotation. In present quantum dot systems, such a combination of operations turns out to be rather difficult due to the strong influence of nuclear spins on the electron spin combined with the rather weak coupling between electron spins and electromagnetic fields. A more direct and robust C-Phase gate realization would therefore be desirable.

New possibilities for constructing efficient two-qubit gates between single-spins are offered~\cite{burkard2} by the recently demonstrated possibility to engineer different Zeeman splittings in neighbouring dots. Using micromagnets placed closed to the quantum dots, up to a 30 mT difference in the local magnetic field in neighbouring dots has been demonstrated~\cite{Pioro,Laird, Obata}, and a 50 mT difference field is realistic~\cite{Tokura}. By local control of the nuclear field, even higher difference fields, up to 1 Tesla, have been been achieved~\cite{Foletti}. Also differences in $g$-factor between neighbouring dots offer similar possibilities and have been realized, both in quantum dots formed in etched GaAs pillars~\cite{Ono} and in InAs nanowires~\cite{Frolov}. The main motivations for these gradients in Zeeman splitting were local addressing of individual spins by spectral selectivity, and the possibility of single-spin rotations by electric-dipole spin resonance~\cite{Tokura}. Strong gradients also permit very fast exchange-controlled single-spin rotations~\cite{coish}.

Here we show that a difference in the local Zeeman splitting $dE_z$ between two quantum dots can be exploited to realize a fast and high-fidelity C-Phase gate under realistic conditions. Taking a broader view, we show that the natural two-qubit gate in tunnel-coupled quantum dots evolves from the exchange gate at $dE_z = 0$ to the C-Phase gate at large $dE_z$. The cross-over depends on the strength of the interdot tunnel coupling and on the magnitude and rise time of the pulse that tilts the double dot potential. We quantify the error probability of the C-Phase gate under realistic conditions, giving insight in the optimal working point for fault-tolerant quantum computation.

At the heart of the C-Phase gate are spin dependent energy shifts, analogous to C-Phase gate realizations with trapped ions~\cite{WinelandRMP} and nuclear spins in molecules~\cite{Vandersypen-Chuang}. As we will show in detail below, in the presence of a gradient, tilting the double dot potential lowers the two-electron states with anti-parallel spins in energy with respect to those with parallel spins. This results in energy shifts $\Delta E_{\uparrow\downarrow}$ and $\Delta E_{\downarrow\uparrow}$ respectively for the two antiparallel spin states. Fast and accurate control of the double dot potential therefore allows adding a controllable spin-dependent phase shift to the anti-parallel spin states, described by the unitary transformation
\begin{equation}
\begin{array}{cccc}
U_{CPhase}= \left(
\begin{array}{cccc}
  1 & 0 & 0 & 0\\
  0 & e^{i\phi_2} & 0 & 0 \\
  0 & 0 & e^{i\phi_1} & 0 \\
  0 & 0 & 0 & 1 \;,
\end{array}
\right)
\end{array}
\label{Eq1}
\end{equation}
expressed in the $\{ |\uparrow\uparrow\rangle$, $|\uparrow\downarrow\rangle$, $|\downarrow\uparrow\rangle$, $|\downarrow\downarrow\rangle \}$ basis and $\phi_{1(2)}=\Delta E_{\uparrow\downarrow(\downarrow\uparrow)}t_{wait}/\hbar$ where $t_{wait}$ is the time the system spent in the tilted position.

For $\phi_1+\phi_2=\pi$, the gate corresponds to a controlled $\pi$-phase gate, up to single-qubit $\hat{z}$ rotations, with phase $\phi_1$ and $\phi_2$ on the first and second qubit respectively. The single-qubit $\hat{z}$ rotations can be left out when the C-Phase is part of a larger sequence of operations, as they can be absorbed into the phase of the subsequent single-spin $\hat{x}$ or $\hat{y}$ rotations~\cite{Vandersypen-Chuang}.

In our analysis, we consider a closed double quantum dot filled with two electrons. The relevant two-electron states are the four states with one $\uparrow$ or $\downarrow$ electron in each dot and the singlet states $S(0,2)$ ($S(2,0)$) with both electrons in the right (left) dot. We also define $S(1,1)=(|\uparrow\downarrow\rangle-|\downarrow\uparrow\rangle)/\sqrt{2}$ and $T_0(1,1)=(|\uparrow\downarrow\rangle+|\downarrow\uparrow\rangle)/\sqrt{2}$. The triplet states with two electrons in the same dot are much higher in energy and can be ignored in the following discussion. The Hamiltonian of this system in the basis $\{ T_+=|\uparrow\uparrow\rangle$, $|\uparrow\downarrow\rangle$, $|\downarrow\uparrow\rangle$, $T_-=|\downarrow\downarrow\rangle$, $S(0,2)$, $S(2,0)\}$ is then given by
\begin{equation}
\begin{array}{cccccc}
H= \left(
\begin{array}{cccccc}
  -E_z & 0 & 0 & 0 & 0 & 0\\
  0 & \frac{-dE_z}{2} & 0 & 0 & t & t\\
  0 & 0 & \frac{dE_z}{2} & 0 & -t & -t\\
  0 & 0 & 0 & E_z & 0 & 0\\
  0 & t & -t & 0 & U-\epsilon & 0\\
  0 & t & -t & 0 & 0 & U+\epsilon\\
\end{array}
\right)
\end{array}
\label{Eq2}
\end{equation}
where $t$ is the interdot tunnel coupling (at the anti-crossing, the $S(1,1)$-$S(0,2)$ separation is $2\sqrt{2}t$), $E_z$ is the average and $dE_z$ the difference in the Zeeman splitting between the two dots, $U$ the energy cost for moving both electrons into the same dot (analogous to the on-site interaction energy in the Hubbard model~\cite{mattis}) and $\epsilon$ the detuning or relative alignment of the potential of the two dots. In this Hamiltonian, we neglect the influence of spin-orbit interaction and magnetic field gradients in the $\hat{x}$ and $\hat{y}$ direction. The effect of the transverse gradients is to couple $T_+$ and $T_-$ to $S(1,1)$, and indirectly (and weakly) to $S(0,2)$ as well. The spin-orbit interaction gives a direct matrix element between the $(1,1)$ triplet states and $S(0,2)$~\cite{danon}. In both cases, the resulting $T_\pm$-$S(0,2)$ anti-crossing will only produce a (small) energy shift and phase evolution of $T_\pm$. Since this phase shift will be different from the phase shift of the anti-parallel spin states, we still obtain a C-Phase gate.

\begin{figure}[!t]
\includegraphics[width=3.4in]{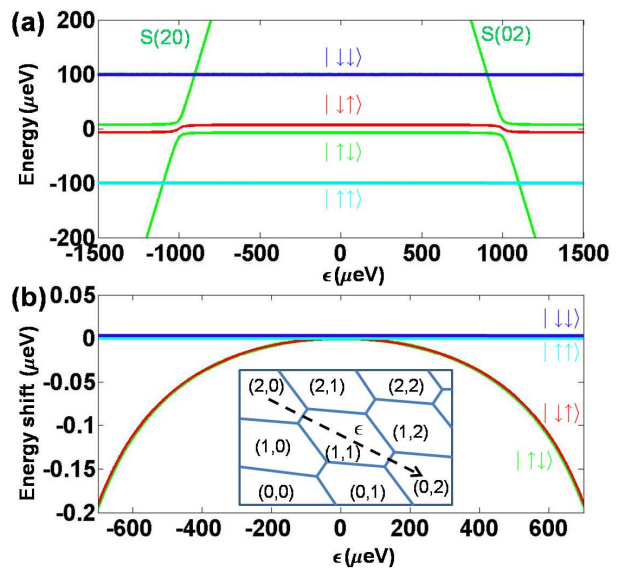}
\caption{(a) Energy level diagram for a double quantum dot containing two electrons, as a function of detuning $\epsilon$ along the $(2,0)$-$(1,1)$-$(0,2)$ axis. The six levels with the lowest energies are shown, and are labelled in the figure. Energies are expressed relative to the $(1,1)$ states at $\epsilon=0$ and $E_z=0$. The parameters used are $U=1000\mu$eV, $t=10\mu$eV, $dE_z=14\mu$eV and $E_z=100\mu$eV.
(b) Energy shift of the levels around $\epsilon=0$ with respect to their energy at $\epsilon=0$.
(Inset) Schematic double-dot charge stability diagram around the $(1,1)$ charge configuration. The dashed line shows the detuning axis along which panel (a) is plotted.}
\label{Fig1}
\end{figure}

The resulting energy level diagram as a function of detuning $\epsilon$ is depicted in Fig.~\ref{Fig1}a. For $\epsilon$ close to zero, the electron spins in the two dots are uncoupled and the eigenstates are $T_+$, $|\uparrow\downarrow\rangle$, $|\downarrow\uparrow\rangle$ and $T_-$. Their energy differences are set by $E_z$ and $dE_z$. The $T_+$ and $T_-$ states don't move in energy with $\epsilon$, since they are not coupled (via tunneling) to S(0,2) or S(2,0). The anti-parallel spin states, in contrast, shift down in energy as $|\epsilon|$ moves away from zero, since they contain a $S(1,1)$ component and the $S(1,1)$ state has an avoided crossing with the $S(0,2)$ and $S(2,0)$ states. By controlling the detuning $\epsilon$ in time, one can add a phase shift between the states with anti-parallel and parallel spins and therefore realize in a single step the C-phase gate presented in Eq.~\ref{Eq1}.

In the following, we will quantify the spin-dependent phase shifts from an effective Hamiltonian acting on the four two-qubit states. Analytical formulas for the phase accumulation $\phi_1$ and $\phi_2$ in Eq.~\ref{Eq1} will help to characterize the speed and fidelity of the C-Phase gate.

\begin{figure}[!t]
\includegraphics[width=3.4in]{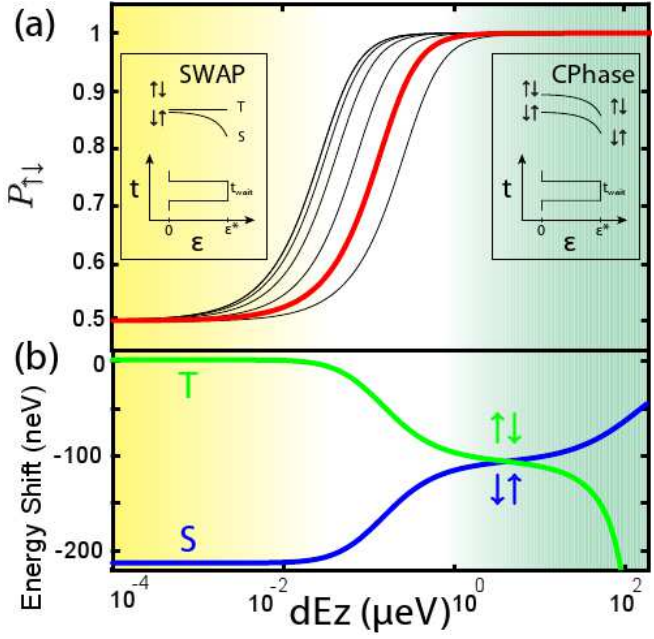}
\caption{(a) Probability of $|\uparrow\downarrow\rangle$ in one of the anti-parallel eigenstates which shows the change in the character of the eigenstates when $dE_z$ is increased, extracted from diagonalisation of $H_e$. Different curves correspond to values of $\epsilon =$ 0, 100, 750, 1200, 1600, 1800 (in red), and 1900 ($U=2000\mu$eV, $t=5\mu$eV). The closer $\epsilon$ is to $U$, the larger the $dEz$ that is needed to change the eigenbasis. (b) Energy shift of the two anti-parallel qubit states as a function of $dE_z$ (extracted from Eq.~\ref{Eq5}) when $\epsilon$ is pulsed from $0$ to $\epsilon^*=1800\mu$eV ($U-\epsilon^*=200\mu$eV, corresponding to the red line in the top panel) . For small $dE_z$, the eigenstates are triplet and singlet and only one state (the singlet) is shifted in energy. At $dE_z=\sqrt2 t$, the energy shift is the same for the two anti-parallel eigenstates.}
\label{Fig2}
\end{figure}

Using a Schrieffer-Wolf transformation and considering $t$ smaller than $U\pm\epsilon \;(>0)$~\cite{SchriefferWolf,PRBLoss}, we can derive an effective Hamiltonian $H_{e}=e^S H e^{-S}$ eliminating at any order of approximation the off-diagonal terms of $H$ between the $(1,1)$ qubit states and $S(0,2)$ and $S(2,0)$. To second order, and expressed in the same basis as $H$ in Eq.~\ref{Eq2}, $S$ has the form
\begin{equation}
\begin{array}{cccccc}
S= \left(
\begin{array}{cccccc}
  0 & 0 & 0 & 0 & 0 & 0\\
  0 & 0 & 0 & 0 & -\gamma(-dE_z) & -\sigma(-dE_z)\\
  0 & 0 & 0 & 0 & \gamma(dE_z) & \sigma(dE_z)\\
  0 & 0 & 0 & 0 & 0 & 0\\
  0 & \gamma(-dE_z) & -\gamma(dE_z) & 0 & 0 & 0\\
  0 & \sigma(-dE_z) & -\sigma(dE_z) & 0 & 0 & 0\\
\end{array}
\right)
\end{array}
\label{Eq3}
\end{equation}
where $\gamma(dE_z)=t/(U-\epsilon-dE_z/2)$ and $\sigma(dE_z)=t/(U+\epsilon-dE_z/2)$. Then, the projection of $H_{e}$ onto the two-qubit subspace $\{|\uparrow\uparrow\rangle$, $|\uparrow\downarrow\rangle$, $|\downarrow\uparrow\rangle$, $|\downarrow\downarrow\rangle\}$ has the form
\begin{equation}
\begin{array}{cccc}
H_{e}= \left(
\begin{array}{cccc}
  -E_z & 0 & 0 & 0\\
  0 & -dE_z/2-\alpha(-dE_z) & \beta(dE_z) & 0 \\
  0 & \beta(dE_z) & dE_z/2-\alpha(dE_z) & 0 \\
  0 & 0 & 0 & E_z \\
\end{array}
\right)
\end{array}
\label{Eq4}
\end{equation}
where
\begin{eqnarray}
\alpha(dE_z,\epsilon) &=& \frac{t^2}{U-\epsilon-dE_z/2}+\frac{t^2}{U+\epsilon-dE_z/2} \\
\beta(dE_z,\epsilon) &=& \frac{\alpha(dE_z,\epsilon)+\alpha(-dE_z,\epsilon)}{2} \;.
\end{eqnarray}
\label{eq5}
In the limit where $\epsilon, dE_z \ll U$, $\alpha$ and $\beta$ are well approximated by $2t^2/U$ and Eq.~\ref{Eq4} reduces to Eq. 37 of Ref.~\cite{burkard2} with exchange interaction $J=4t^2/U$.

When $dE_z$ is non-zero but small with respect to $t$, the energy eigenstates are $\{|\uparrow\downarrow\rangle, |\downarrow\uparrow\rangle\}$ close to $\epsilon=0$ but $\{S(1,1),T_0(1,1)\}$ for larger $\epsilon$. This can be seen in Fig.~\ref{Fig2}a, where the $\uparrow\downarrow$ component of one of the anti-parallel eigenstates is shown as a function of $dE_z$ for 7 values of $\epsilon$. The exchange gate takes advantage of this condition by pulsing $\epsilon$ non-adiabatically from zero to a finite value~\cite{PettaScience}. Then constructing a C-Phase gate requires combining two exchange gates with a single-spin rotation.

When $dE_z$ is large, the magnetic field gradient is the dominant part of the Hamiltonian, hence $\{|\uparrow\downarrow\rangle$,$|\downarrow\uparrow\rangle\}$ are eigenstates as long as $\epsilon$ is smaller than U, as seen also in Fig.~\ref{Fig2}a. In this case, only the energies of the antiparallel states change. The exchange gate~\cite{PettaScience}is then no longer possible and the corresponding eigenenergies vary with $\epsilon$, as seen from Fig.~\ref{Fig2}b and from the following analytical expressions:
\begin{eqnarray}
E_{\uparrow\downarrow/\downarrow\uparrow}(dE_z,\epsilon)=-\frac{\alpha(dE_z)+\alpha(-dE_z)}{2} \mp A(dE_z)
\label{Eq5}\\
A(dE_z,\epsilon)=1/2\sqrt{dE_z^2+dE_z\Sigma(dE_z,\epsilon)+\Pi(dE_z,\epsilon)}
\end{eqnarray}
\label{eq6}

where $\Sigma(dE_z,\epsilon)=2(\alpha(-dE_z,\epsilon)-\alpha(dE_z,\epsilon))$ and $\Pi(dE_z,\epsilon)=2(\alpha(dE_z,\epsilon)^2+\alpha(-dE_z,\epsilon)^2)$.

\begin{figure}[!t]
\includegraphics[width=3.4in]{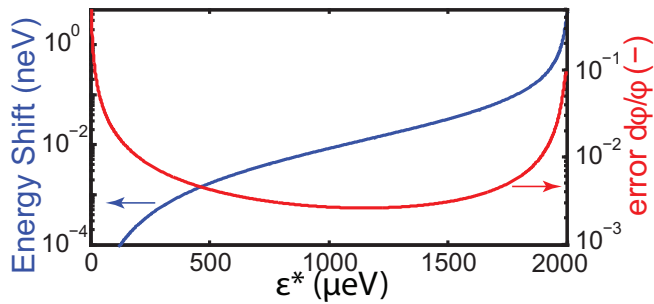}
\caption{Expected energy shift with respect to the case $\epsilon=0$ as a function of detuning $\epsilon^*$ ($t=dEz/\sqrt{2}=5\mu$eV and $U=2000\mu$eV). The closer to the $S(1,1)$-$S(0,2)$ anti-crossing, the faster the gate. Also shown is the relative errors in the phase $\phi$ due to $1 \mu$eV fluctuations in $\epsilon$. Between $\epsilon^*=300\mu$eV and $\epsilon^*=1850\mu$eV, errors are below the threshold for fault-tolerant quantum computation.}
\label{Fig3}
\end{figure}

This means that a C-Phase gate can be realized in a single step via a fast pulse in $\epsilon$, from $\epsilon=0$ to $\epsilon=\epsilon^*$. The closer to the $S(1,1)$-$S(0,2)$ crossing the accumulation of phase takes place, the larger the energy shift hence the faster the operation is. We note that when $dE_z\approx\sqrt2 t$, $A=dE_z$ to second order, so $|\uparrow\downarrow\rangle$ and $|\downarrow\uparrow\rangle$ acquire the same energy difference $\Delta E_{\uparrow\downarrow(\downarrow\uparrow)}=E_{\uparrow\downarrow(\downarrow\uparrow)}(dE_z,\epsilon^*)-E_{\uparrow\downarrow(\downarrow\uparrow)}(dE_z,0)$ and thus accumulate the same phase when $\epsilon$ is pulsed. In this case, the C-Phase gate is obtained for $\phi_1 = \phi_2 =\pi/2$ (as noted before, the general condition for C-Phase is $\phi_1 + \phi_2 = \pi$). A ten ns gate duration, requires an energy shift of $\sim 0.1$ $\mu$eV, which is in reach with $dE_z=5 \mu$eV as seen from Fig.~\ref{Fig3} (we note that phase evolution during the finite rise time of the pulses can be easily included in the timing calibration). Gate durations of 1 ns and less are possible by using stronger tunnel couplings or larger pulse amplitudes. If $dE_z$ is made larger too, this will not affect the gate operation. These gate durations are $10^4$-$10^5$ times shorter than reported spin-echo times in GaAs~\cite{T2long}, so error probabilities due to decoherence will be below the $10^{-4}$ accuracy threshold for fault-tolerant quantum computation~\cite{Nielsen-Chuang}.

Finally, we analyze for realistic conditions the gate errors that can be expected from other sources than decoherence, with the help of the analytical expressions obtained above. To simplify the analysis, we consider the case where $\phi_1=\phi_2=\phi$. As in the case of the exchange gate~\cite{PettaScience}, $\phi$ will be sensitive to potential fluctuations due to charge noise or gate voltage noise since we allow a small amount of mixing between charge and spin states to obtain the phase shift. The effect of potential fluctuations manifests itself predominantly as fluctuations in $\epsilon$, rather than as fluctuations in $t$ or U~\cite{fujisawa}. The closer to the $S(1,1)$-$S(0,2)$ crossing the operating point is located, the more sensitive to detuning fluctuations the C-Phase gate is, as can be seen from Fig.~\ref{Fig3}. From Eq.~\ref{Eq5}, analytical expressions can be derived for $d\phi/\phi$ induced by detuning fluctuations and the results are also shown in Fig.~\ref{Fig3}, taking a realistic value of $1\mu$eV detuning fluctuations~\cite{hayashi}. We see that relative errors $d\phi/\phi$ down to $6\times 10^{-3}$ can be obtained for the 10 ns C-Phase gate, which translate to a gate error probability $(d\phi/\phi)^2=4 \times 10^{-5}$.

A second possible source of errors is the random value of the (quasi-static) nuclear field in the two dots. This affects the time-evolution during C-Phase in two ways. First, it causes a dephasing that is independent of $\epsilon$. This dephasing has been shown to be (largely) reversible by spin-echo techniques and since C-Phase commutes with dephasing, it can be simply embedded in existing spin-echo sequences. Second, there is an additional contribution to the $\epsilon$-dependent energy shift $\phi$, but this is only a second order effect in the $dE_z$ fluctuations. The slow variation of $\phi$ with $dE_z$ is seen in Fig.~\ref{Fig2}, in particular as $dE_z$ approaches $t$. Gate error probabilities due to $dE_z$ fluctuations can be evaluated from Eq.~\ref{Eq5} in the regime of fast gate operation where $dE_z \ll U-\epsilon^* \ll U$ and, for $dE_z=\sqrt2 t$, are approximately given by $d\phi/\phi=\delta/(U-\epsilon)$ where $\delta$ is the size of the fluctuations in $dE_z$ ($\approx 20$ neV in GaAs). Therefore, for the values used above to obtain a 10 ns gate, the gate error probability $(d\phi/\phi)^2$ due to the nuclear field randomness would be only $\approx 10^{-8}$.

The last source of error we consider is possible non-adiabaticities that may occur if the rise time of the detuning pulses is too short. This is relevant when the anti-parallel eigenstates are slightly different between the two pulse positions $\epsilon=0$ and $\epsilon=\epsilon^*$. For the parameters used above, we quantify this difference via the square of the overlap between the eigenstates at the two pulse positions and it is different from $1$ only by $8 \times 10^{-4}$. For instantaneous detuning pulses, the resulting gate error probability calculated from simulations is of the same order of magnitude, $\sim 2 \times 10^{-3}$. However, for realistic rise times in the range of 0.3-1 ns and again using the same $t$ and $\epsilon$ as above, the gate error probability is well below $10^{-4}$. By using shaped pulses, the adiabaticity can be improved even further~\cite{requist}.

In conclusion, we present and analyze a one-step implementation of a C-Phase gate between two single-spin qubits in a double quantum dot. This gate is the natural replacement of the exchange gate when large magnetic field or $g$-factor gradients are present. We estimate that gate operation times of a few nanoseconds can be achieved under realistic conditions with fidelities above the accuracy threshold for fault-tolerance. To probe the spin-dependent phase shifts produced by the C-Phase gate, a Ramsey type experiment can be used, with the C-Phase gate in between two single-spin rotations. We expect that this work will impact ongoing experiments aimed at the realization of elementary quantum gates and simple quantum protocols with single-spin qubits in a variety of quantum dot systems.

\begin{acknowledgments}

We thank G. Burkard, S. Florens, S. Frolov, H. Keijzers, D. Loss, K. Nowack and L. Schreiber for useful discussions and the Foundation for Fundamental Research on Matter (FOM), the European Research Council and a Marie Curie fellowship for financial support.

\end{acknowledgments}




\end{document}